%% file: jet_base.tex
\documentclass[useAMS,usenatbib]{mn2e}
\usepackage{times}
\usepackage{url}
\usepackage{graphicx}
\graphicspath{{./figs/}}

\newcommand\Fi{\ensuremath{F_\mathrm{i}}}
\newcommand\ri{\ensuremath{r}}
\newcommand\dr{\ensuremath{\Delta r}}
\newcommand\di{\ensuremath{\delta_\mathrm{i}}}
\newcommand\de{\ensuremath{\delta}}
\newcommand\rg{\ensuremath{r_\mathrm{g}}}
\newcommand\relDelta{\ensuremath{r^2-2r+a^2 }}

\newcommand\ii{\ensuremath{I_\mathrm{i}}}

\newcommand\gi{\ensuremath{g_\mathrm{lp}}}

\newcommand\hb{\ensuremath{h_\mathrm{base}}}

\newcommand\uh{u_\mathrm{h}}
\newcommand\gaphi{\gamma^{(\phi)}}
\newcommand\dif[1]{\mathrm{d}#1}

\newcommand\accel{\ensuremath{\mathcal{A}}}

\newcommand\jhb{\ensuremath{h_\mathrm{base}}}
\newcommand\jht{\ensuremath{h_\mathrm{top}}}

\newcommand\pli{\ensuremath{\Gamma}}

\title[Irradiation of an Accretion Disc by a Jet]{Irradiation of an
  Accretion Disc by a Jet: General Properties and Implications for
  Spin Measurements of Black Holes}

\author[T. Dauser et al.]{T.\ Dauser$^{1}$\thanks{E-mail:
    thomas.dauser@sternwarte.uni-erlangen.de}, J. Garcia$^{2}$,
  J. Wilms$^{1}$, M. B\"ock$^{1,3}$, L. W. Brenneman$^{4}$, 
  M. Falanga$^{5}$, \newauthor K. Fukumura$^{6}$, 
  and  C. S. Reynolds$^{2}$ \\
  $^{1}$ Dr.\ Karl
  Remeis-Observatory and Erlangen Centre for Astroparticle Physics,
  Sternwartstr.~7, 96049 Bamberg, Germany \\
  $^{2}$ Department of Astronomy and Maryland Astronomy Center for
  Theory and Computation, University of Maryland, College Park, MD
  20742, USA\\
  $^{3}$ Max-Planck-Institut f\"ur Radioastronomie, Auf dem H\"ugel 69, 
  53121 Bonn, Germany \\
  $^{4}$ Harvard-Smithsonian Center for Astrophysics, 60 Garden
  Street, Cambridge, MA 02138, USA \\
  $^{5}$ International Space Science Institute, Hallerstrasse 6, 3012, 
  Bern, Switzerland \\
  $^{6}$ Astrophysics Science Division, NASA Goddard Space Flight Center,
  Code 663, Greenbelt, MD 20771, USA}
\begin{document}

\pagerange{\pageref{firstpage}--\pageref{lastpage}} \pubyear{2012}

\maketitle
\label{firstpage}

\begin{abstract}
  X-ray irradiation of the accretion disc leads to strong reflection
  features, which are then broadened and distorted by relativistic
  effects. We present a detailed, general relativistic approach to
  model this irradiation for different geometries of the primary X-ray
  source. These geometries include the standard point source on the
  rotational axis as well as more jet-like sources, which are radially
  elongated and accelerating. Incorporating this code in the
  \textsc{relline} model for relativistic line emission, the line
  shape for any configuration can be predicted. We study how different
  irradiation geometries affect the determination of the spin of the
  black hole. Broad emission lines are produced only for compact
  irradiating sources situated close to the black hole. This is the
  only case where the black hole spin can be unambiguously
  determined. In all other cases the line shape is narrower, which
  could either be explained by a low spin or an elongated source. We
  conclude that for those cases and independent of the quality of the
  data, no unique solution for the spin exists and therefore only a
  lower limit of the spin value can be given.
\end{abstract}


\section{Introduction}
\label{sec:introduction}

Due to the vicinity of the X-ray emitting region in Active Galactic
Nuclei (AGN) and X-ray binaries to the central compact object, it is
expected that the observed X-ray spectrum will show signs of
relativistic effects \citep{Fabian1989}. Such effects were first seen
in the skew symmetric shape of the fluorescent Fe K$\alpha$ line from
these objects \citep[][and references
  therein]{tanaka1995a,reynolds2003a}.  These relativistic lines are
present in a significant fraction of AGN spectra
\citep{guainazzi:06a,nandra2007a,Longinotti2008a,Patrick2011a} and
Galactic black hole binaries
\citep{Miller2007,Duro2011a,Fabian2012b}. Recent work has been
applying a more self-consistent approach by including models for the
relativistic distortion of the full reflection spectrum during the
data analysis \citep[see,
  e.g.,][]{Zoghbi2010a,Duro2011a,Fabian2012a,Dauser2012a}.

The relativistic distortion of the reflection spectrum is determined
by the spin of the black hole, $a$, the geometry of the reflector, and
our viewing direction on the system, parametrised through an
appropriate inclination angle, $i$. Additionally a primary source of
radiation has to exist in order to produce the observed reflection and
also the underlying continuum. Initial assumptions were that the
primary source consists of a hot corona around the inner regions of
the disc, as Comptonization of soft disc photons in such a corona
naturally produces a power law spectrum which fits the observations
\citep{Haardt1993a,Dove1997a}. Under the assumption that the intensity
of the hard radiation scattered back onto the disc by the corona is
proportional to the local disc emissivity, the irradiation of the
accretion disc would be $I(r) \propto r^{-3}$ for the outer parts, and
gradually flatten towards the inner edge of the disc for a standard
\citet{Shakura1973a} disc.

With the advent of high signal to noise data from satellites such as
\textsl{XMM-Newton}, however, measurements showed a disagreement with
the Fe K$\alpha$ line profiles predicted by this coronal geometry. For
many sources, the data favoured disc emissivities that are much
steeper in the inner parts of the accretion disc \citep[see,
e.g.,][]{wilms2001a,Miller2002b,fabian2002a,Fabian2004a,Fabian2012b,Brenneman2006a,Ponti2010a,Brenneman2011a,Gallo2011a,Dauser2012a}. Variability
studies of the broad iron lines pose additional problems for standard
corona models. In such studies the time variability of the continuum
flux, i.e., the primary hard X-ray radiation, is compared to the flux
in the lines, which are produced by the reflected radiation. This
allows to probe the connection between the primary and the reflected
radiation. In a coronal geometry, one expects a positive correlation
between the strength of the relativistically distorted reflection
spectrum and the primary continuum \citep{Martocchia1996a}. This is in
contrast to what is observed: Measurements of \mbox{MCG$-$6-30-15}
\citep{Fabian2003b,Miniutti2003a} revealed large variations of the
direct radiation, while the reflected component remained constant.

As shown by \citet{Martocchia2002b}, \citet{Fabian2003b},
\citet{Miniutti2003a}, and \citet{Vaughan2004a} for the case of
\mbox{MCG$-$6-30-15}, in a geometry in which the illuminating
continuum is assumed to be emitted from a source on the rotational
axis at height $h$ above the black hole, strong light bending yields
properties of the reflected radiation that are consistent with the
observations.  Figure~\ref{fig:lamp_post_scheme} illustrates this
``lamp post'' geometry \citep{Matt1991a,Martocchia1996a}. In general,
data and predicted line shapes show very good agreement
\citep[see][]{Wilkins2011a,Duro2011a,Dauser2012a}. The lamp post model
also explains the observed connection between the luminosity and the
reflection strength: For a primary source very close to the black
hole, most of the photons are focused on the accretion disc, producing
a strong reflection component. Therefore less photons are left over to
contribute to the continuum component, which is directly emitted
towards the observer \citep{Miniutti2004a}. For an increasing height
of the hard X-ray source this effect gets weaker and thus more photons
can escape, which strengthens the continuum radiation and, depending
on the flux state of the X-ray source, weakens the reflected flux
\citep{Miniutti2004a,Miniutti2006a}.

Based on the earlier work on the lamp post geometry presented above,
\citet{Fukumura2007a} provided a more detailed treatment of the
emissivity for arbitrary spin and anisotropic emission of the primary
source. Off-axis sources were first investigated by
\citet{Ruszkowski2000a}. Using similar methods \citet{Wilkins2012a}
presented a ray tracing method working on Graphics Processing Units
(GPUs), which can calculate irradiation profiles for almost arbitrary
geometries of the primary sources, now also including sources extended
along and perpendicular to the rotational axis. In this \emph{paper}
we present a complete method to derive irradiation profiles in the
lamp post geometry (Sect.~\ref{sec:theory}), including radially
extended and accelerating sources
(Sect.~\ref{sec:an-extended-ray}). Using this formalism, we introduce
an implementation of the lamp post geometry as a fitting model for
relativistic reflection, which can be applied to more realistic
expectations of theoretical jet models (Sect.~\ref{sec:relline}). In
Sect.~\ref{sec:influence-lamp-post} we analyse the shape of the
reflection features predicted by the different types of irradiating
sources. In particular we concentrate on the implications for spin
measurements, the different assumptions for the geometry of the
primary source have (Sect.~\ref{sec:impic-meas-spin}). These results
are quantified by simulating such observations for current instruments
(Sect.~\ref{sec:simul-observ-with}). Finally, we summarise the main
results of the paper in Sect.~\ref{sec:summary}.

\section{Theory}\label{sec:theory}

\subsection{Introduction}
We calculate the shape of the relativistic line in the lamp-post
geometry by following the radiation emitted from the primary source on
the axis of symmetry of the accreting system to the accretion disc and
from there to the observer. Due to the deep gravitational potential
close to the black hole, the photon trajectories are bent and their
energies red-shifted. Moreover the relativistic movement of the
accretion disc alters the energy flux incident on the disc and the
shape of the observed line through the relativistic Doppler
effect. Using techniques introduced by \citet{Cunningham1975}, the
flux seen from a certain element of the accretion disc under a
specific inclination can be predicted \citep[e.g., ][]{Speith1995} and
summed up to the complete spectrum of the source. This problem has
been extensively studied in the past and there are several models
available to predict the relativistic smearing given a certain
emissivity of the accretion disc \citep[e.g.,
][]{Fabian1989,Laor1991,Dovciak2004a,Brenneman2006a,Dauser2010a}. In
these models the intensity emitted from the accretion disc (the
so-called ``emissivity'') is parametrised as a power law
$r^{-\epsilon}$ with index $\epsilon$, where $r$ is the distance to
the black hole. The standard behaviour is $\epsilon=3$, which is
proportional to the energy release in a standard \citet{Shakura1973a}
disc.

This emissivity can also be calculated directly from an irradiating
source, the so-called ``primary source''. In the following we will use
a source situated on the rotational axis of the black hole for this
purpose. By applying the same ray-tracing techniques used to trace
photons from the disc to the observer, we can also calculate the
proper irradiation of the accretion disc by the primary source. The
radial dependency of this irradiation is equal to the reflected
radiation, i.e., the emissivity, which was previously modelled by a
power law. In this paper we concentrate on the \emph{irradiation} of
the accretion disc. In order to be able to compare our results to
observational data, we also need to calculate the ray tracing from the
accretion disc to the observer, which is done with the
\textsc{relline}-code \citep{Dauser2010a}.

\subsection{Photon Trajectories in the Lamp Post Geometry}

\begin{figure}
  \includegraphics[width=\columnwidth]{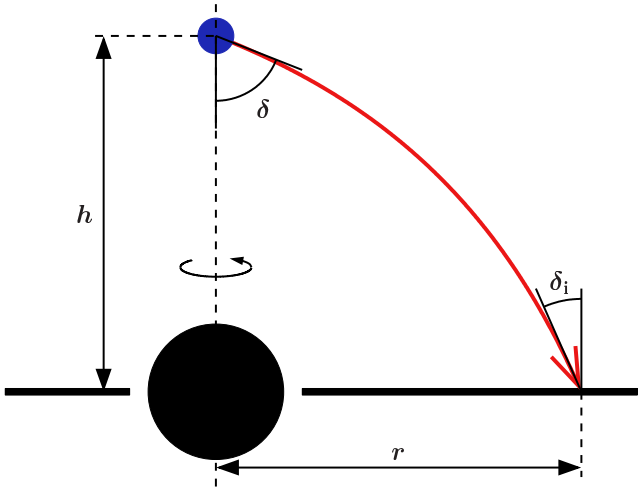}
  \caption{\label{fig:lamp_post_scheme} A schematic drawing of the
    components in the lamp post geometry. The primary source of
    photons (blue) is situated above the rotational axis of the black
    hole and is emitting photons (red), which hit the accretion
    disc. }  
\end{figure}
In the following we will concentrate on a simplified geometry by
assuming a point-like, photon emitting primary source at a height $h$
above the rotational axis of the black hole. This source irradiates a
thin, but optically thick accretion disc
(Fig.~\ref{fig:lamp_post_scheme}). Relativistic photon trajectories in
the lamp post geometry were first investigated by \citet{Matt1991a}
and used by \citet{Martocchia1996a} in order to explain the very large
equivalent width of the iron K$\alpha$ line in some AGN. A more
detailed discussion of effects in this geometry was presented by
\citet{Martocchia2000a}, including a discussion of the influence of
the black hole's spin on the overall spectra. 

As a good physical explanation of this hard X-ray source on the
rotation axis is the base of a jet \citep{Markoff2004a}, we call this
geometry also the ``jet base geometry''. This interpretation is highly
supported by the earlier work of \citet{Ghisellini2004a}, where it is
shown that all AGN are capable of forming jets. This is achieved by
inventing the concept of ``aborted'' jets for radio-quite quasars and
Seyferts, which are produced when the velocity of the outflowing
material is smaller than the escape speed. Therefore such a jet
extends only a small distance from the black hole, producing only a
negligible amount of radio flux while at the same time it strongly
irradiates the inner accretion disc in X-rays, which produces the
observed, highly relativistic reflection \citep{Ghisellini2004a}. This
interpretation is encouraged by works showing that direct and
reprocessed emission from such a jet base is equally capable in
describing the observed X-ray spectrum as a corona above the accretion
disc and also yields a self-consistent explanation of the full radio
through X-ray spectrum of many compact sources
\citep{Markoff2005a,Maitra2009a}. In addition a direct connection
between the X-rays and the radio can explain the correlation between
observed radio and X-ray flares of Microquasars such as
\mbox{GX~339$-$4} \citep{Corbel2000a} or \mbox{Cyg~X-1}
\citep{Wilms2007a}. We note that a similar kind of connection is also
indicated in some measurements of AGN like \mbox{3C~120}
\citep{Marscher2002a} or \mbox{3C~111} \citep{Tombesi2012a}.
Additionally, evidence of a direct influence of the jet on the black
hole in Microquasars is growing recently. Namely, \citet{Narayan2012a}
observe a direct correlation between the jet power and the spin value
from analysing a small sample of sources.

In the following we will briefly summarise the most important
equations required for deriving the photon trajectories from a source
on the rotational axis of the black hole. As we are dealing with
potentially rapidly rotating black holes, we choose the
\citet{Kerr1963} metric in \citet{Boyer1967} coordinates to describe
the photon trajectories. Following, e.g., \citet{Bardeen1972}, the
general photon momentum is given by
\begin{equation}\label{eq:ph_mom}
  p_t = -E, \;\;
  p_r = \pm E \sqrt{V_r}/\Delta, \;\;
  p_\theta = \pm E\sqrt{V_\theta}, \;\;
  p_\phi = E\lambda, 
\end{equation}
with
\begin{eqnarray}
  \Delta &=& r^2 - 2r + a^2\\
  V_r &=& (r^2+a^2)^2 - \Delta(q^2+a^2)\\
  V_\theta &=& q^2 - \cos^2\theta \left[ \frac{\lambda}{\sin^2\theta} -
    a^2  \right] 
\end{eqnarray}
for a certain distance $r$ and spin $a$ of the black hole. The spin is
defined in such a way that its absolute value ranges from \mbox{$0 \le
  |a| \le 1$}, where at the maximal value, $a_\mathrm{max}=1$, the
event horizon would have to rotate at the speed of
light.\footnote{\citet{1974Thorne} showed that the realistic upper
  limit of the spin is more likely $a_\mathrm{max} = 0.998$.}  Note
that all equations in this paper are given in units of $G \equiv M
\equiv c \equiv 1$. The different signs in equation (\ref{eq:ph_mom})
are for increasing (upper sign) and decreasing (lower sign) values of
$r$ and $\theta$. Here the conserved quantities are the total energy,
$E$, the angular momentum parallel to the rotational axis of the black
hole, $\lambda$, and the \citet{Carter1968} constant, $q$. The latter
is given by
\begin{equation}\label{eq:carter_q}
  q =  \sin\de  \sqrt{ \frac{h^2 - 2h + a^2}{ h^2 + a^2} } \quad.
\end{equation}
Assuming the source to be located on the symmetry axis of the system
($\theta=0$) simplifies the calculation significantly, as $p_\theta$
does only take a real value if $\lambda = 0$, and therefore equation
(\ref{eq:ph_mom}) simplifies to
\begin{equation}\label{eq:ph_mom_simple}
 p_\mu = E (-1,\;\; \pm\sqrt{V_r}/\Delta,\;\; \pm E |q| ,\;\; 0) \quad.
\end{equation}
Then the trajectory can finally be calculated numerically from the
integral equation
\begin{equation} \label{eq:eq_motion_lamp}
  \int_h^r \frac{\dif{r'}}{\pm\sqrt{V_{r'}}} = 
    \int_0^\theta \frac{\dif{\theta'}}{\pm\sqrt{V_{\theta'}}} \quad.
\end{equation}
\citep[following ][]{Chandrasekhar1983,Speith1995}.  Note that the
sign changes at turning points of the photon trajectory.  In this case
the left part of equation (\ref{eq:eq_motion_lamp}) has to be split
into two integrals, each going from and to the turning point,
respectively. Note that no turning points in the $\theta$ direction
have to be taken into account, as photons which initially fly towards
the disc, will not exhibit a turning point before crossing the
equatorial plane and thus before hitting the accretion disc.

\subsection{Illumination of the Accretion Disc}
\label{sec:intens-accr-disc}

In order to calculate the incident intensity on the accretion disc, we
first have to consider the geometric effects intrinsic to the lamp
post setup.  Without any relativistic effects the intensity impinging
on the accretion disc for an isotropic primary emitter is given by
\begin{equation}\label{eq:intens_flat}
  \ii (r,h) \propto \frac{\cos\di}{r^2+h^2} =
  \frac{h}{\left(r^2+h^2\right)^{\frac{3}{2}}} \quad.
\end{equation}
This means that already for flat space time the irradiated intensity
strongly depends on the radius.

Due to the strong gravity the photon trajectories will be
significantly bent, i.e., in our setup the photons will be ``focused''
onto the inner regions of the accretion disc
(Fig.~\ref{fig:lamp_post_scheme}), modifying the radial intensity
profile. Note that this focusing depends on both, the height, $h$, and
the initial direction of the photon, parametrised by the angle
between the system's axis of symmetry and the initial direction of the
photon, $\delta$.

Using the equations of the previous section we developed a ray-tracing
code using similar techniques as those presented in
\citet{Dauser2010a}.  With this code we are able to calculate photon
trajectories from the point of emission $(h,\de)$ at the primary
source to the accretion disc, yielding the location $(r,\di)$ where
this specific photon hits the disc. As the primary source is located
on the rotational axis of the black hole, the trajectory of the photon
is uniquely determined by the $q$-parameter (equation
\ref{eq:carter_q}). The incident point is then calculated by solving
the integral equation (\ref{eq:eq_motion_lamp}) for $r$ in the case of
$\theta=\mathrm{\pi}/2$.

Knowing where the isotropically emitted photons hit the accretion
disc, we can derive the photon flux incident on its surface. As the
photons are designed to be emitted at equally spaced angles \de, the
distance \dr\ between these points is related to the incident
intensity. Photons emitted in $[\de,\de+\Delta\de]$ are distributed on
a ring on the accretion disc with an area of $A(r,\dr)$. The proper
area of such a ring at radius $r$ with thickness \dr\ is given by
\begin{equation}
A(r,\dr) = 2\mathrm{\pi} r \cdot \sqrt{\frac{r^4+a^2r^2+2a^2r}{r^2-2r+a^2}} \dr
\end{equation}
in the observer's frame of rest \citep{Wilkins2012a}. In order to
calculate the irradiation in the rest frame of the accretion disc, we
have to take into account its rotation at relativistic speed. The area
of the ring will therefore be contracted. Using the Keplerian
velocity profile deduced from the Kerr metric \citep{Bardeen1972}, the
disc's Lorentz factor is
\begin{equation}\label{eq:gamma}
  \gaphi = 
  \frac{\sqrt{\relDelta}(r^{3/2} + a)}
       {r^{1/4} \sqrt{r\sqrt{r} + 2a -3\sqrt{r}}
        \sqrt{r^3 + a^2r+2a^2} }
\end{equation}
\citep[][see also \citealt{Wilkins2011a,Wilkins2012a}]{Bardeen1972}.
Taking into account that the photons are emitted at equally spaced
angles, we finally find that for isotropic emission the geometric
contribution to the incident intensity has to be
\begin{equation}
  \ii^{\mathrm{geo}} = \frac{\sin\de}{A(r,\dr)\gaphi} \quad.
\end{equation}
Because of the relative motion of the emitter and the accretion disc,
as well as because of general relativistic effects, the irradiated
spectrum will be shifted in energy \citep{Fukumura2007a}. Using the
initial four-momentum at the primary source
\begin{equation}
  u^\mu_\mathrm{h} = (u^t_\mathrm{h},0,0,0)
\end{equation}
and the corresponding four-momentum on the accretion disc 
\begin{equation}
  u^\mu_\mathrm{d} = u^t_\mathrm{d}(1,0,0,\Omega)
\end{equation}
together with the photon's momentum (equation \ref{eq:ph_mom_simple}),
the energy shift is
\begin{equation}\label{eq:energy-shift}
  \gi = \frac{E_\mathrm{i}}{E_\mathrm{e}} = \frac{p_\mu
    u_\mathrm{d}^\mu} {p_\nu u_\mathrm{h}^\nu} =
  \frac{\left(\ri\sqrt{\ri}+a\right)\sqrt{h^2-2h+a^2}}
       {\sqrt{\ri}\sqrt{\ri^2-3\ri+2a\sqrt{\ri}}\sqrt{h^2+a^2}} 
\end{equation}
The components of the four-velocities are calculated from the
normalising condition $u_\mu u^\mu = -1$ \citep[see, e.g.,
][]{Bardeen1972}.

As the number of photons is conserved we can write
\begin{equation}
  N^{(\mathrm{ph})}_\mathrm{e} \Delta t_\mathrm{e} \Delta E_\mathrm{e} = 
    \mathrm{const.} = 
  N^{(\mathrm{ph})}_\mathrm{i} \Delta t_\mathrm{i} \Delta E_\mathrm{i}  
    \quad ,
\end{equation}
where $N^{(\mathrm{ph})}_\mathrm{e}$ ($N^{(\mathrm{ph})}_\mathrm{i}$) is the
emitted (incident) photon flux. Assuming a power law shape of the
emitted radiation
\begin{equation}
  N^\mathrm{ph}_\mathrm{e}  = E_\mathrm{e}^{-\Gamma} \quad,
\end{equation}
the photon flux on the accretion disc is given by
\begin{equation}
  N^\mathrm{ph}_\mathrm{i} (\ri, a)  = E_\mathrm{i}^{-\Gamma}\cdot 
  \gi(\ri, a)^{\Gamma} \quad,
\end{equation}
as $\Delta E_\mathrm{e}/\Delta E_\mathrm{i} = 1/\gi$ and $\Delta
t_\mathrm{e}/\Delta t_\mathrm{i} = \gi$. Due to the relativistic
energy shift, the incident photon flux now also depends on where the
photon hits the accretion disc ($\ri$) and which spin the black hole
has.  Using this result, we can finally calculate the incident flux on
the accretion disc
\begin{equation} \label{eq:intens_on_ad}
  \Fi(\ri, h ) =  \ii^{\mathrm{geo}} \cdot \gi^{\pli}    = 
  \frac{\sin\de \gi^{\pli} }{A(r,\dr)\gaphi} \quad.
\end{equation}
This is in line with the results obtained by
\citet{Fukumura2007a}\footnote{Note that \citet{Fukumura2007a} use the
  spectral index $\alpha$, whereas we use the photon index
  $\Gamma$. Both quantities are related by $\Gamma = \alpha +1$.}.

\begin{figure}
  \includegraphics[width=\columnwidth]{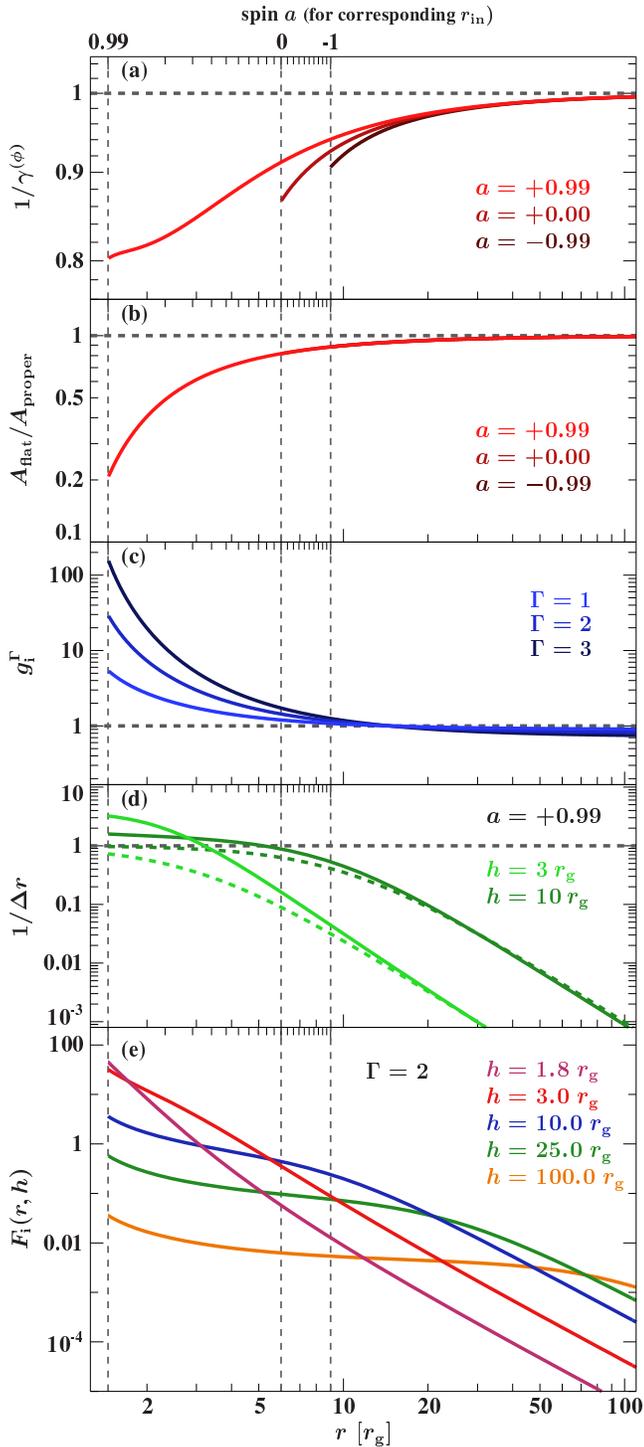}  
  \caption{\label{fig:simple_gamma_eshift}Relativistic factors which
    influence the incident flux on the accretion disc compared to the
    emitted intensity at the primary source (see equation
    \ref{eq:intens_on_ad}). If not stated in the figure explicitly we
    use $a=0.99$ and assume that the primary source is an isotropic
    emitter. The vertical dashed lines indicate the location of the
    innermost stable circular orbit (ISCO) for certain values of
    spin. (a)~The inverse beaming factor (red), which determines the
    influence of length contraction (see equation \ref{eq:gamma}) on
    the incident flux. (b)~The impact of the proper
    area. (c)~Geometric intensity distribution on the accretion disc
    for the relativistic (solid) and Newtonian case
    (dashed). (d)~Energy shift, which the photon experiences when
    travelling from the primary source to the accretion disc, taken to
    the power of $\Gamma$ (blue). (e)~Combined irradiating flux on the
    accretion disc for a primary source at different heights but equal
    luminosity. }
\end{figure}
In order to understand the influence of the different relativistic
parameters on the incident intensity,
Fig.~\ref{fig:simple_gamma_eshift} shows the single components of
equation (\ref{eq:intens_on_ad}). A similar discussion of these
components is also given by \citet{Wilkins2012a}. We assume that the
primary source is an isotropic emitter. All effects are strongest for
small radii and will therefore be most important for high spin, where
the accretion disc extends to very low radii. First, length
contraction reduces the area of the ring as seen from the primary
source. In the rest frame of the accretion disc, this ``contraction''
implies an effectively larger area and therefore the incident flux
decreases with increasing $v^\phi$ proportional to the inverse Lorentz
factor $1/\gaphi$ (Fig.~\ref{fig:simple_gamma_eshift}a). When compared
to flat space time, the area of disc close to the black hole is
additionally enhanced in the Kerr metric
(Fig.~\ref{fig:simple_gamma_eshift}b).  Interestingly, this effect is
almost independent of the spin of the black hole\footnote{deviations
  are less than 0.2\%}. However, compared to the effect induced by the
energy shift (equation \ref{eq:energy-shift}), the change in area is
only a minor effect.  Depending on the power law index, $\Gamma$, the
energy shift of the photons hitting the disc is the strongest factor
influencing the reflection spectrum. For a source on the rotational
axis of the black hole the change in irradiated flux can be as large
as a factor of 100 (Fig.~\ref{fig:simple_gamma_eshift}c), depending
strongly on the steepness of the primary spectrum. This amplification
factor depends on the height of the emitting source (it becomes larger
for increasing height) and decreases for larger radii. Especially for
low $h$ light bending focuses photons towards the disc and
additionally enhances the irradiation of the inner regions. The dashed
lines in Fig.~\ref{fig:simple_gamma_eshift}d show how $1/\dr$
decreases in flat space just due to geometrical reasons following
equation (\ref{eq:intens_flat}). The fully relativistic treatment
(solid lines) reveals a focusing of the photons towards the black
hole. But compared to the ``effectively enhanced'' irradiation of the
inner parts due to the energy shift
(Fig.~\ref{fig:simple_gamma_eshift}c), the relativistic focusing is
only a minor effect. In summary, the power law index, $\Gamma$, has
the strongest influence on the irradiation profile at small radii,
while the height of the emitting source mostly affects the outer parts
of the disc.

Finally, Fig.~\ref{fig:simple_gamma_eshift}e combines all effects and
shows the incident flux in the rest frame of the accretion disc. In
general this plot confirms our overall understanding of the lamp post
geometry: Sources at low height strongly irradiate the inner parts but
almost not the outer parts. For an increasing height of the source
more and more photons hit the outer parts of the accretion disc and an
increasing region of more constant irradiation at roughly $h/2$ is
created. In order to check our simulation for consistency, a thorough
check against the calculations of \citet{Fukumura2007a} was done and
we could validate the result from Fig.~\ref{fig:simple_gamma_eshift}e
at high precision. The same is true for the stationary point source
solution of \citet{Wilkins2012a}.

\subsection{Emissivity Profiles in the Lamp Post Geometry}
\label{sec:emiss-prof-lamp}

Since for a simple accretion disc the local disc emissivity is roughly
$\propto r^{-3}$, in the description of observations it is common to
parametrise the disc emissivity profile through
\begin{equation}\label{eq:emis_index_def}
  F(r,h) \propto r^{-\epsilon}
\end{equation}
where $\epsilon$ is called the emissivity index. Note that in this
representation the information of the normalisation of the emissivity
profile is lost. But as usually the luminosity of the irradiating
source is not known, this is not very important for reflection
studies. Our calculations easily allow to determine the radius
dependent emissivity index and thus phrase our results in a language
that is directly comparable with observations.

\begin{figure}
  \includegraphics[width=\columnwidth]{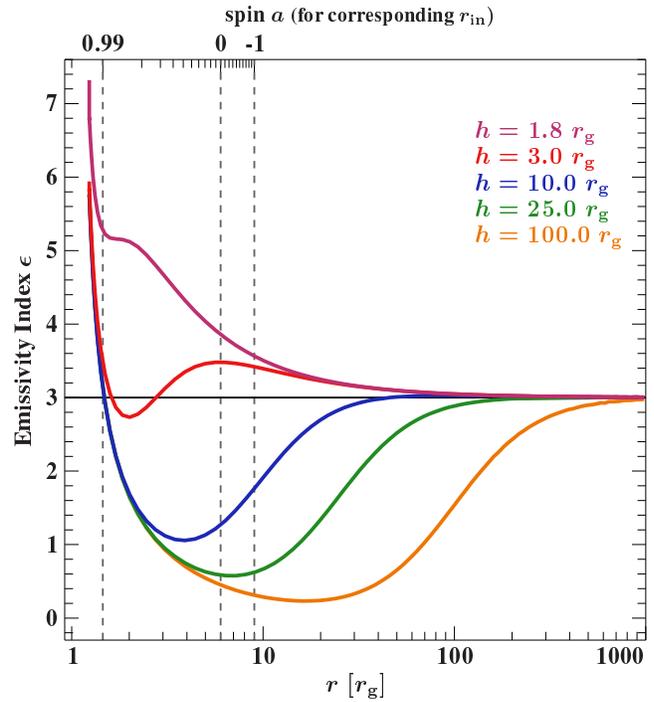}
  \caption{\label{fig:jb_emis_intens} The emissivity index $\epsilon$
    as defined in equation (\ref{eq:emis_index_def}) of the radiation
    irradiating the accretion disc from a primary source at different
    heights $h$. }
\end{figure}
Figure~\ref{fig:jb_emis_intens} shows the emissivity profile for
different heights of the primary source. Regardless of the specific
height, three different radial zones are visible in all profiles.
Firstly, for large radii the index converges always towards its value
in flat space ($\epsilon = 3$, see equation \ref{eq:intens_flat}). The
closer the emitting source is to the black hole, the faster $\epsilon$
converges towards this value.

The zone closer than $2\,\rg$ from the black hole is characterised by
a strong steepening of the emissivity profile towards the inner edge
of the accretion disc. Except for an extremely low height, this
steepening is almost independent of the height of the primary
source\footnote{Clearly, the absolute flux for a certain luminosity is
  highest for a low emitter (see Fig.~\ref{fig:simple_gamma_eshift}e),
  but we usually do not measure the absolute intensity, but only the
  emissivity index.}.  Hence, a large emissivity at low radii, which
is usually interpreted as ``strong focusing of a low height emitter'',
is not directly related to the height of the primary source. Instead,
the steepness almost solely depends on the relativistic boosting of
the primary photons and especially on the steepness $\Gamma$ of the
primary spectrum (see Fig.~\ref{fig:simple_gamma_eshift}c).

\begin{figure}
  \includegraphics[width=\columnwidth]{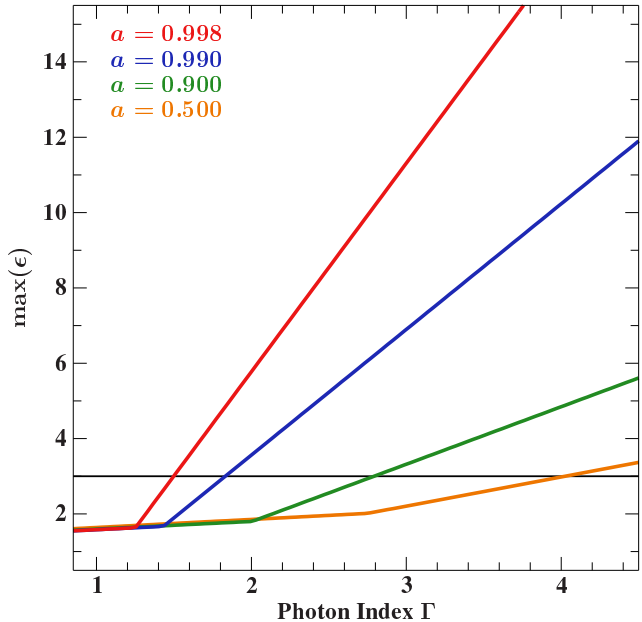}
  \caption{The maximum possible emissivity index at the inner regions
    of the disc ($r < 10\,\rg$) for a certain photon index
    $\Gamma$. The height of the source was chosen to be
    $h=10\,\rg$. However, note that we showed in
    Fig.~\ref{fig:jb_emis_intens} that for $h>3\,\rg$ the steepening
    at the inner edge of the disc is almost independent of the height.}
  \label{fig:max_emis_index}
\end{figure}

As has been mentioned in the introduction
(Sect.~\ref{sec:introduction}), many sources are observed to have very
steep emissivity indices, i.e., values of \mbox{$\epsilon=5$-$10$} are
normal. Similar to the emissivity profile in
Fig.~\ref{fig:jb_emis_intens}, we can also derive the maximal possible
emissivity index for a certain value of spin and steepness of the
input spectrum. For a standard lamp post source at a height of at
least 3\,\rg, this information is plotted in
Fig.~\ref{fig:max_emis_index}.
In this case, no large emissivities are expected if the black hole is
not a maximal rotator, i.e., roughly for $a < 0.9$. If the input
spectrum is hard ($\Gamma < 2.5$), the emissivity index is flatter
than the standard index of $\epsilon=3$.  Even for highly rotating
black holes the actual emissivity at the inner edge of the accretion
disc is only steep if the incident spectrum is very soft ($\Gamma >
2.5$).

If we want to apply the information of Fig.~\ref{fig:max_emis_index}
to a certain observation, we have to take into account that the
emissivity is generally parametrised in form of a broken power law. In
detail this means that below a certain ``break radius''
($r_\mathrm{br}$) one single emissivity index is used to describe the
steep emissivity and above that it is usually fixed to the canonical
$r^{-3}$ behaviour. Generally, break radii are found to be in the
range of 3\,\rg\ \citep[e.g., in \mbox{1H0707$-$495}, ][]{Dauser2012a}
up to 6\,\rg\ \citep[e.g., in \mbox{NGC~3783},
][]{Brenneman2011a}.\footnote{Note that the value of the break radius
  is highly correlated with the emissivity index when trying to
  constrain both by observation.} Hence, the emissivity index we
measure in observations accounts for the average steepness in the
range of $r_\mathrm{in}$ to $r_\mathrm{br}$ and will therefore be
lower than the maximal emissivity. We note that the emissivity indices
found in many observations are close to or above the maximal allowed
emissivity index as defined in Fig.~\ref{fig:max_emis_index}. For
example $\epsilon\approx 5$ for $\Gamma=1.8$ in \mbox{NGC~3783}
\citep{Brenneman2011a}, $\epsilon\approx 6$ for $\Gamma=2$ in
\mbox{MCG$-$6-30-13} \citep{Brenneman2006a}, $\epsilon>6.8$ for
$\Gamma=1.37$ in \mbox{Cygnus~X-1} \citep{Fabian2012b}, or $\epsilon
\approx 10$ for $\Gamma=3.3$ in \mbox{1H0707$-$495}
\citep{Dauser2012a}. As the emissivity index obtained from these
measurements is averaged over the innermost few \rg\, these
emissivities can therefore not be properly explained solely by the
lamp post geometry. Despite these issues, however, in some cases it is
possible to use Fig.~\ref{fig:max_emis_index} to decide if a measured
value for the emissivity index can reasonably be explained in the lamp
post geometry. This method has been successfully applied by
\citet{Duro2011a} to find a unique and consistent solution to describe
the reflection spectrum of Cyg~X-1.

\subsection{The incident angle}
\label{sec:incident-angle}

\begin{figure*}
  \includegraphics[width=\textwidth]{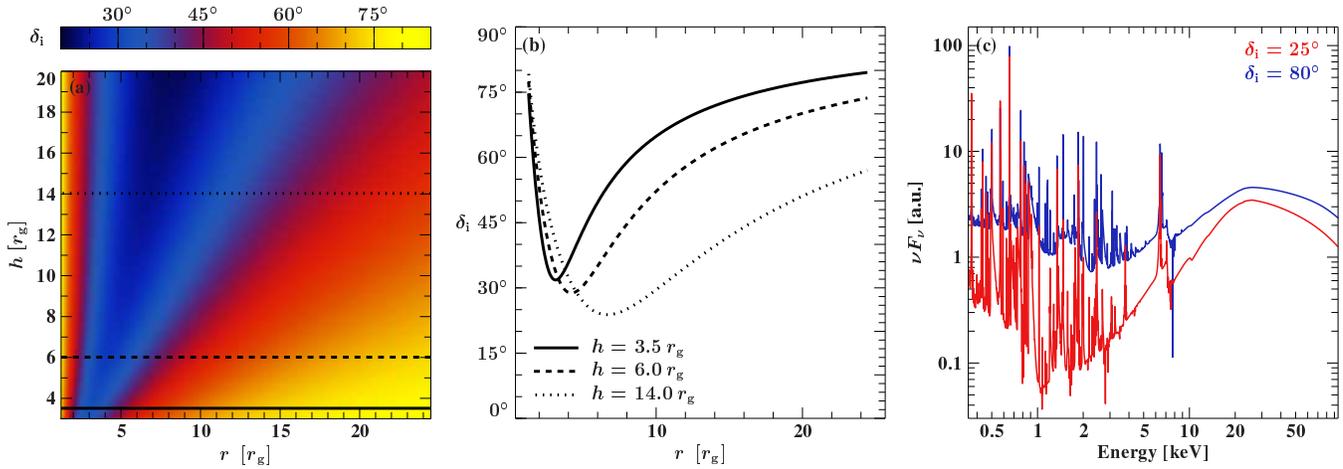}
  \caption{\label{fig:incident_angle} (a) 2d image showing the
    incident angle \di\ of photons on the accretion disc. The spin of
    the black hole is $a=0.998$. (b) incident angle for the three
    lines (solid, dashed, dotted) marked in subfigure (a). (c) Sample
    reflection spectra for different incident angles $\di =
    25^\circ$ (red) and $\,80^\circ$ (blue)
    calculated with the \textsc{xillver} code
    \citep{Garcia2010a,Garcia2011a}. }
\end{figure*}

The incident angle $\delta_\mathrm{i}$ of the irradiated radiation is
important for modelling the reflected spectrum, as it determines the
typical interaction depth of the reflected photon and therefore
strongly influences the limb-darkening of the reflected radiation
\citep[e.g.,][]{Svoboda2009}. Constructing a normal vector on the
disc, $\delta_\mathrm{i}$ is given by
\begin{equation}\label{eq:q-delta-connection}
  \cos\delta_\mathrm{i} = \frac{p_\perp}{|p|} = \frac{(p_d)_\mu
    \left(n_d^{(\theta)}\right)^\mu } {(p_d)_\nu (u_d)^\nu}
  \Bigg|_{\theta=\mathrm{\pi}/2} = \frac{q}{\ri u_d^t(\ri, a)} \quad .
\end{equation}
Figures~\ref{fig:incident_angle}a and~b show $\delta_\mathrm{i}$ for
different heights of the primary source and assuming an isotropic
primary emitter. For most disc radii the photons hit the disc at a
shallow angle, except for a small fraction of disc. The location and
width of the steeper in-falling photons depends on the height of the
primary source.

The effect of the incidence angle of the illumination in the reflected
spectrum from an accretion disc has been discussed in
\citet{Garcia2010a}.  In the calculation of reflection models, the
boundary at the surface of the disc is defined by specifying the
intensity of the radiation field that illuminates the atmosphere at a
particular angle. Using their equation (19) and (37), this can be expressed
as
\begin{equation}\label{eqinc}
I_\mathrm{inc} = \left( \frac{2n}{4\mathrm{\pi}} \right) \frac{\xi}{\cos\di},
\end{equation}
where $n$ is the gas density (usually held fixed), and $\xi$ is the
ionisation parameter that characterises a particular reflection model
\citep{Tarter1969a}. Consequently, for a given ionisation parameter,
varying the incidence angle varies the intensity of the radiation
incident at the surface. This has interesting effects on the
ionisation balance calculations. If the photons reach the disc at a
normal angle ($\di=0$), the intensity has its minimum value, but the
radiation can penetrate into deeper regions of the atmosphere
producing more heating. On the contrary, for grazing incidence
($\di=90$) $I_\mathrm{inc}$ increases, resulting in a hotter
atmosphere near the surface; but the radiation field thermalizes at
smaller optical depths, which yields lower temperature in the deeper
regions of the disc. Evidently, these changes in the ionisation
structure will also affect the reflected spectrum (see
Fig.~\ref{fig:incident_angle}c).  The narrow component of the emission
lines are expected to be emitted relatively near the surface, where
photons can easily escape without being absorbed or scattered. On the
other hand, the broad component of the emission lines, and in
particular the ones from high $Z$ elements such as iron, are produced
at larger optical depths ($\tau\sim 1$), and therefore are more likely
to be affected by changes in the ionisation structure of the slab.

However, \citet{Garcia2010a} showed that in a general sense, reflected
spectra resulting from models with large incidence angles tend to
resemble models with higher illumination. This means that the changes
introduced by the incidence angle can be mimicked by correcting the
ionisation parameter to account for the difference introduced in the
illumination. The current analysis shows that below $7\,\rg$ the
incidence angle can vary as much as $25-80$ degrees, equivalent to a
change in the ionisation parameter by more than a factor of~5.
Figure~\ref{fig:incident_angle}c shows the reflected spectra for these
two incidence angles predicted by the \textsc{xillver} code
\citep{Garcia2010a,Garcia2011a} using the same ionisation parameter.
The effect of the incidence angle is evident.

\section{An extended Ray-Tracing Code}
\label{sec:an-extended-ray}
So far we assumed that the emitting primary source is at rest with
respect to the black hole. If the primary emitter is the base of the
jet, however, then it is far more likely that the primary source is
moving. Typical speeds at the jet base can already be relativistic
\citep{McKinney2006a}. Furthermore, if the irradiating source is a
jet, then we need to relax our assumption of a point-like emitter and
include the radial extent of the jet. As we show in this section by
taking into account both of these extensions, the line shape is
significantly affected.

\subsection{A Moving Jet Base}
\label{sec:moving-jet-base}

First investigations of a moving source irradiating the accretion disc
and its implication for the Fe K$\alpha$ were done by
\citet{Reynolds1997a}. \citet{Beloborodov1999a} investigated the
coupling between the moving primary source and the reflected
radiation. Using general relativistic ray tracing techniques,
\citet{Fukumura2007a} and \citet{Wilkins2012a} calculated the
illumination profile on the accretion disc.

We first assume that the emitting primary point source is moving at a
constant velocity $\beta = v/c$. The most prominent effect of a moving
jet base compared to a jet base at rest is the Doppler boosting of
radiation in the direction of the moving blob, i.e., away from the
accretion disc. This boosting also means that the energy shift of the
photon between primary source and accretion disc changes with
velocity. 
The 4-velocity is then given by
\begin{equation} \label{eq:4velo_moving}
  \uh^\mu = \uh^t 
  \left(1, \frac{\dif{r}}{\dif{t}}, 0 ,0 \right)
\end{equation}
The velocity $\beta$ as seen by an observer at the same location in
the locally non-rotating frame (LNRF) is connected to $\dif{r}/\dif{t}$
through
\begin{equation}
  \beta = \frac{e_\mu^{(r)} \uh^\mu }{e_\nu^{(t)} \uh^\nu }
  = \frac{r^2 + a^2}{\Delta}\cdot \frac{\dif{r}}{\dif{t}} \quad,
\end{equation}
where $e_\mu^{(\nu)}$ are the tetrad basis vectors for $\nu = t, r,
\theta, \phi$ \citep[see ][]{Bardeen1972}.

In order to calculate the trajectory of a photon emitted at an angle
$\delta'$ from the moving source, we transform from the moving frame
to the stationary, locally non-rotating frame at the same
location. This means that the photon is emitted at an angle $\de$ in
the stationary system according to
\begin{equation}
  \cos\de = \frac{\cos\delta' - \beta}{1 - \beta\cos\delta'} \quad,
\end{equation}
depending on the velocity of the source. Following, e.g.,
\citet{Krolik1999}, it can be easily shown that this approach implies
that the intensity observed in the stationary frame will be altered by
a factor of
\begin{equation}
  \mathcal{D}^{2} \quad,
\end{equation}
where $\mathcal{D}$ is the special relativistic Doppler factor, which is
defined in our case as
\begin{equation}
  \mathcal{D} = \frac{1}{\gamma(1+\beta\cos\de)} \quad, 
\end{equation}
with $\gamma = 1/\sqrt{1 - \beta^2}$ being the Lorentz factor of the
moving source. Using the transformed intensity, we can now can apply
the stationary calculations from Sect.~\ref{sec:intens-accr-disc} to
the new emission angle $\de$ to obtain the irradiation of the
accretion disk by a moving source. 

Additionally, we have to calculate the proper energy shift between the
accretion disc and the moving source, with the 4-velocity given by
equation~(\ref{eq:4velo_moving}). Similarly to
equation~(\ref{eq:energy-shift}) this energy shift is given by
\begin{equation} \label{eq:gi_v}
  \gi =  \frac{(p_\mathrm{d})_\mu
    u_\mathrm{d}^\mu} {(p_\mathrm{h})_\nu u_\mathrm{h}^\nu} =
  \frac{\gi(\beta=0)}{\gamma\left(1 \mp \frac{\sqrt{(h^2+a^2)^2 - \Delta(q^2+a^2)}}
  {h^2 +    a^2} \beta \right)}
\; .
\end{equation}
Note that for large heights $\gi$ simplifies to its special
relativistic limit, $\gi = \mathcal{D} \gi(\beta=0)$.  Using the
results we already obtained for a stationary primary source (see
Sect.~\ref{sec:intens-accr-disc}), we can finally write down the total
flux the accretion disc sees from a moving source:
\begin{equation} \label{eq:intens_on_ad_velocity} \Fi(\ri, h, \beta )
  = \frac{\mathcal{D}^2 \Fi(\ri, h, \beta=0 ) } {\left[ \gamma\left(1
        \mp \frac{\sqrt{(h^2+a^2)^2 - \Delta(q^2+a^2)}} {h^2 + a^2}
        \beta \right) \right]^{\pli}} \quad.
\end{equation}

\begin{figure*}
  \centering
  \includegraphics[width=\textwidth]{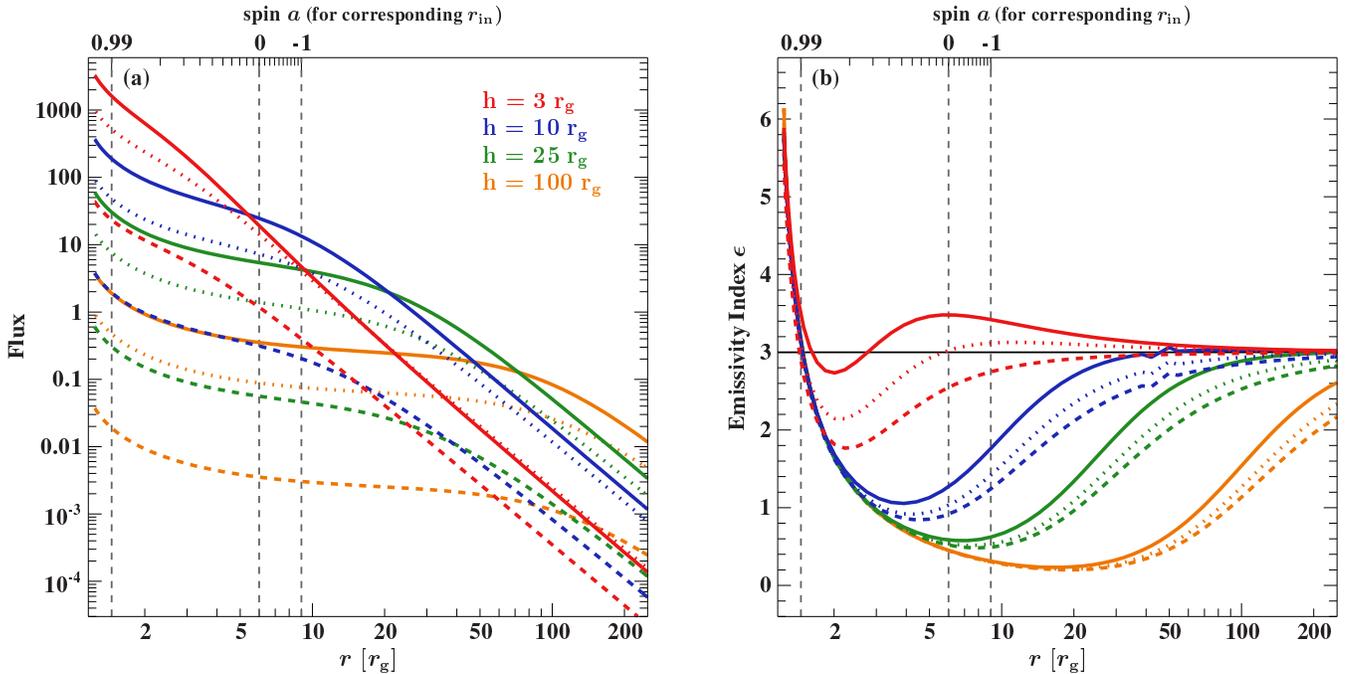}
  \caption{\label{fig:beta_emis_profiles} (a) Irradiating flux and
    (b) Emissivity profiles similar to Fig.~\ref{fig:jb_emis_intens},
    which show the impact of a moving jet base with velocities $v =
    0\,c$ (solid), $v=0.5\,c$ (dotted), and $v=0.9\,c$ (dashed).}
\end{figure*}

Figure~\ref{fig:beta_emis_profiles} shows the dependency of the
emissivity index on the velocity of the jet base.
In general the irradiating flux decreases significantly with
increasing speed of the jet base, as the photons are boosted away from
the black hole (Fig.~\ref{fig:beta_emis_profiles}a).

Comparing the shape of the emissivity profiles for a moving jet base
(Fig.~\ref{fig:beta_emis_profiles}b), the intermediate region of the
accretion disc ($3$--$100\,\rg$) experiences an increase of
irradiation with increasing velocity of the jet base. On the other
hand the emissivity profile of the very inner regions of the accretion
disc does not depend on the movement of the jet. For high spin ($a >
0.9$) the accretion disc extends down to these very small radii and
due to the steep emissivity most of the reflected radiation comes from
there. If the accretion disc only extends down to $6\,\rg$, as is the
case for $a=0$, the irradiation of the innermost regions can differ
almost up to a factor of 2 in emissivity index depending on the
velocity of the jet base.

\subsection{Irradiation by an Elongated Jet}
\label{sec:irrad-an-extend}

\begin{figure*}
  \centering
  \includegraphics[width=\textwidth]{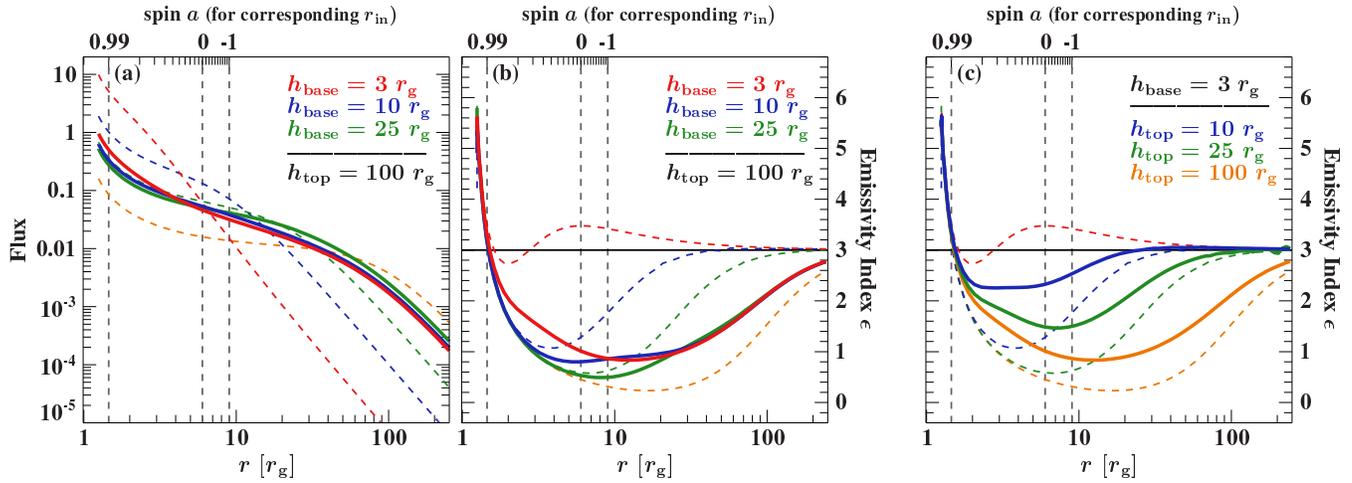}
  \caption{\label{fig:integ_height_emis} Emissivity profiles similar
    to Fig.~\ref{fig:jb_emis_intens}, which show the impact of an
    elongated jet compared to a point source. For comparison the
    dashed lines show the emissivity of a point-like emitting source
    with same colours as in Fig.~\ref{fig:jb_emis_intens}, i.e., red,
    blue, green, and orange for 3\,\rg, 10\,\rg, 25\,\rg, and
    100\,\rg, respectively. For varying the base of the emitting region
    the irradiating flux (a) and the emissivity profile (b) is
    shown. In (c) the top height of the jet is altered. }
\end{figure*}

It is straightforward to extend the previous discussion of a moving
jet base to the the case of an extended jet \cite[see also
][]{Wilkins2012a}. We simply describe the incident radiation for an
elongated jet by many emitting points at different heights, weighted
by the distance between these points. Emissivity profiles for the
extended jet are shown in Fig.~\ref{fig:integ_height_emis}.
In general, the shape of the emissivity profile in the case of an
extended source does not differ significantly from that of a
point-like source. Similar to the moving jet base (see
Sect.~\ref{sec:moving-jet-base}), the irradiation of the inner regions
of the accretion disc ($r < 2 \rg$) only differs in normalisation but
not in shape (Fig.~\ref{fig:integ_height_emis}). However, the regions
of the disc that are a little bit further outwards ($> 3\rg$) are
affected at a much greater fraction by extending the emission
region. But despite these large differences, the overall shape of the
emissivity profile does not change, i.e., that the general properties
analysed in Sect.~\ref{sec:emiss-prof-lamp} are still valid in the
case of elongated jets.

The influence of changing the location of the jet base while fixing
the top height is depicted in Fig.~\ref{fig:integ_height_emis}a,b.
Interestingly the emissivity profile is not very sensitive to the
location of the jet base. On the other hand, fixing the jet base at a
low value (3\,\rg) and then increasing the radial extent of the jet
(Fig.~\ref{fig:integ_height_emis}c) strongly alters the emissivity
profile. The more the top is away from the base, the larger the
deviations become compared to the profile of the jet base (red, dashed
line) and the more the irradiation resembles the one from the upper
part of the jet. Comparing the extended profiles to the ones for a
point-like primary source (dashed lines) reveals that the extended
emission creates an irradiation pattern that could have similarly been
produced by a point source at an intermediate height in between \jhb\
and \jht, too. This implies that if we measure an emissivity profile
similar to one for a low source height
(Fig.~\ref{fig:integ_height_emis}, dashed red line), the jet cannot be
extended or there is no significant amount of radiation from the upper
parts irradiating the disc. One way to explain the lack of photons is
that the jet does not have a uniform velocity, but the jet base is at
rest and from there on the particles are very efficiently accelerated
\citep[see ][]{McKinney2006a} such that the radiation is beamed away
from the accretion disc.

\subsection{Jet with Constant Acceleration}\label{sec:jet-with-constant}
 
\begin{figure*}
  \centering
  \includegraphics[width=\textwidth]{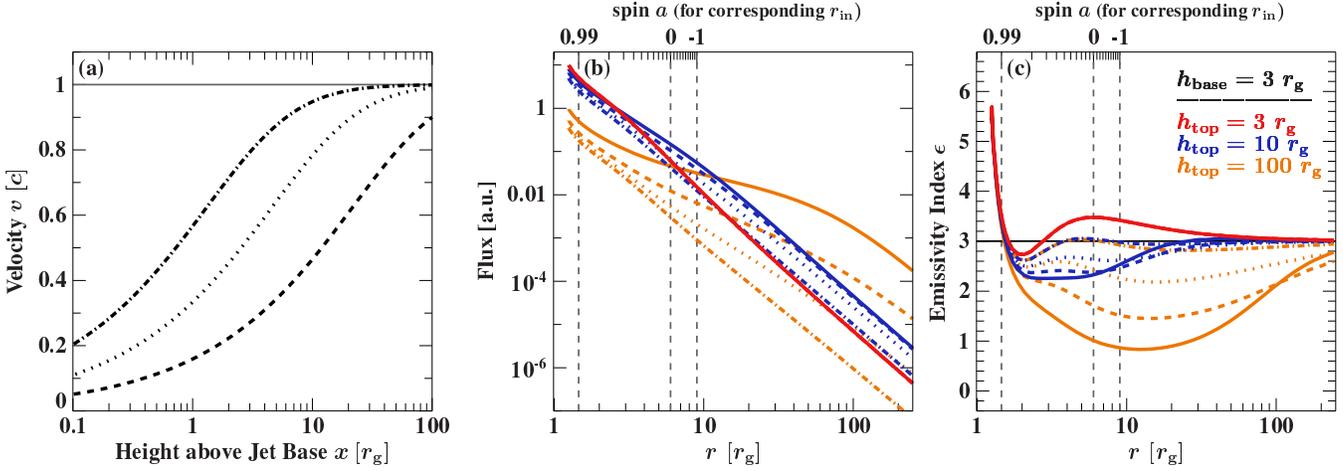}
  \caption{\label{fig:accel_emis} (a) Velocity of the jet assuming a
    constant acceleration and the jet base at rest. The spin of the
    black hole is $a=0.998$, and the power law index $\Gamma=2$. The
    acceleration is parametrised by specifying the velocity $v_{100}$
    the jet has at $100\,\rg$ above the jet base.  Lines are plotted
    for $v_{100}=0.9\,$c (dashed), $v_{100}=0.99\,$c (dotted), and
    $v_{100}=0.999\,$c (dashed-dotted). (b) and (c) Emissivity
    profiles for an extended jet with the constant acceleration shown
    in (a). Parameters are the same as in a). The solid line displays
    the emissivity profile for a stationary jet. Despite the different
    heights, each jet is assumed to have an equal luminosity. }
\end{figure*}

Having analysed the effect of a moving primary source and the profile
of an extended jet, we are able to combine these effects to form a
more realistic approach. It is likely that the actual base of the jet
is stationary or has at least a velocity normal to the disc plane that
is much less than the speed of light. Above the jet base the particles
are efficiently accelerated \citep{McKinney2006a} to higher energies
and in the end to very fast velocities seen, e.g., in Very Long
Baseline Interferometry (VLBI)
measurements \citep[see, e.g., ][]{Cohen2007a}. In the following we
will assume the simplest case by using a constant acceleration
$\accel$ of the particles. In this case the velocity evolves as
\citep[see, e.g.,][]{Castillo2006a}
\begin{equation}\label{eq:v1}
  \beta(t) = \frac{\accel t}{\sqrt{ c^2 + \accel^2t^2}} \quad,
\end{equation}
where the time $t$ is given by
\begin{equation}
  t = \sqrt{\frac{x^2}{c^2} + 2 \frac{x}{\accel}}
\end{equation}
for $x = h-\hb$. Therefore the acceleration to reach a specific
velocity $\beta$ at height $h$ is given by
\begin{equation}
  \accel = \frac{  \gamma - 1 }{h-\hb}\quad,
\end{equation}
where $\gamma$ is the Lorentz factor. Figure~\ref{fig:accel_emis}a
displays the velocity (equation \ref{eq:v1}) inside the jet for constant
acceleration.

The irradiation of the accretion disc in this setup is shown in
Fig.~\ref{fig:accel_emis}b. The effect of the accelerated movement
shows up at larger radii by steepening the emissivity compared to the
stationary jet and the for all heights the profile gets more similar
to a point source at the jet base. This result confirms the general
picture that for a large acceleration the accretion disc sees only the
lowest part, as most of the upper part of the jet is strongly beamed
away from the disc. This means that if we measure a localised, low
height of the emitting source, it could also be the base of a strongly
accelerating jet. Additionally, Fig.~\ref{fig:accel_emis}c reveals
that the stronger the acceleration, the more the emissivity profiles
resembles the canonical $r^{-3}$ case for all but the innermost radii
($r>2\,\rg$).

\section{Discussion}
\label{sec:discussion}

\subsection{\textsc{relline\_lp} --- A new relativistic line model}
\label{sec:relline}

Using the approach mentioned in Sect.~\ref{sec:intens-accr-disc}, the
lamp post geometry was incorporated in the \textsc{relline} model
\citep{Dauser2010a}. This model was designed to be used with common
data analysis tools such as \textsc{xspec} \citep{Arnaud1996a} or
\textsc{isis} \citep{Houck2000a} for modelling relativistic
reflection. It can either predict a single line shape or it can be
used as a convolution model smearing a complete ionised spectrum
\citep[such as the \textsc{reflionx} model][]{ross2007a}. The new
\textsc{relline\_lp} model can be downloaded from
\url{http://www.sternwarte.uni-erlangen.de/research/relline/}.

The calculations in the lamp post geometry are used to determine the
proper irradiation profile and replace the artificial broken power law
emissivity in the \textsc{relline} model. Besides the standard point
source in the lamp post geometry, we also included the extended
geometries presented in Sect.~\ref{sec:an-extended-ray}, i.e.,
elongated and moving primary sources. As the information of the ray
tracing is tabulated, the \textsc{relline\_lp} model is evaluated very
quickly and thus well suited for data modelling.

\subsection{Influence of the lamp post parameters on the shape of the
  reflection features}
\label{sec:influence-lamp-post}

\begin{figure*}
  \centering
  \includegraphics[width=\textwidth]{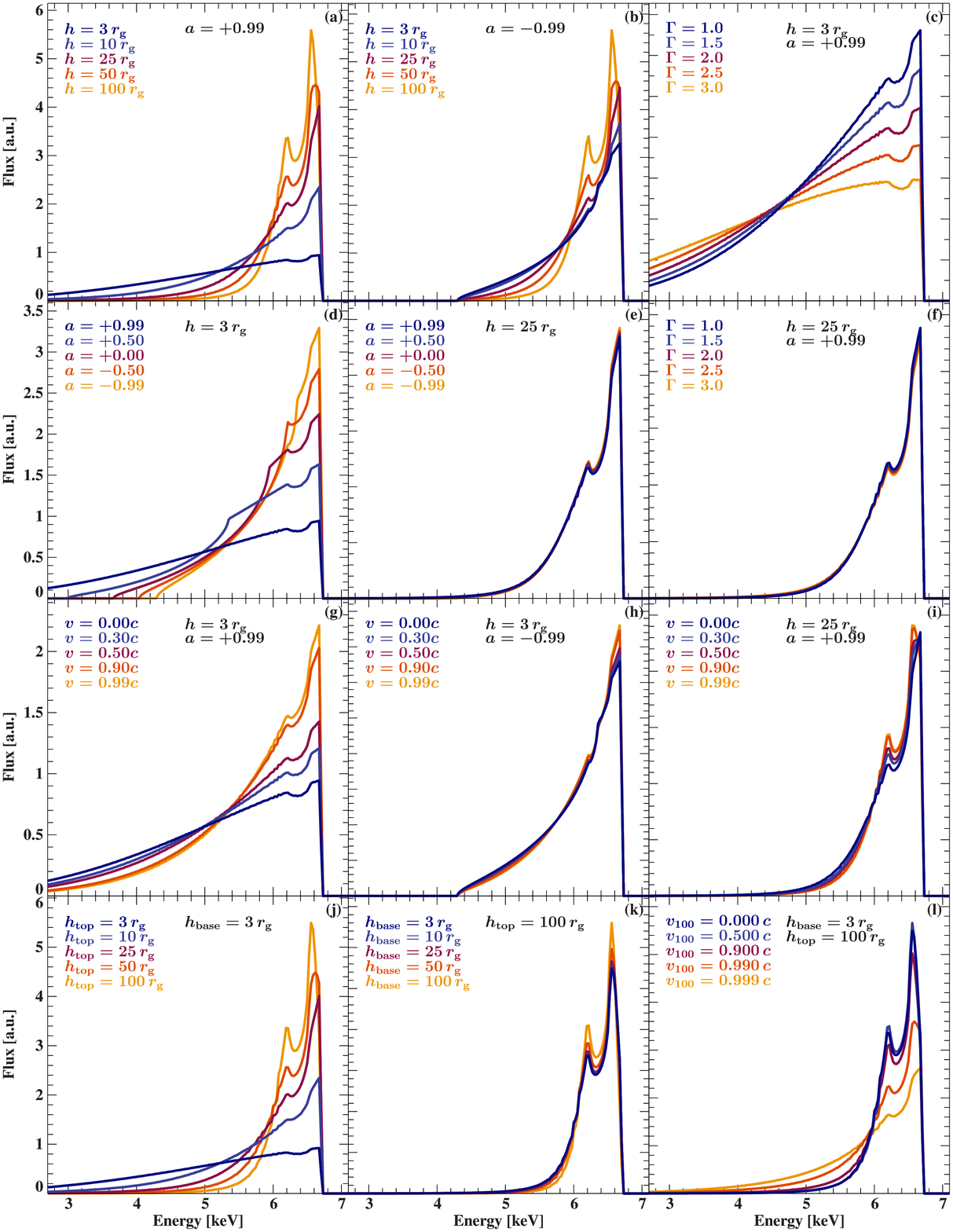}
  \caption{\label{fig:lines_relline_lp} Line Profiles of the
    \textsc{relline\_lp} model for different parameters. If not stated
    differently we assume $a = 0.99$ and $h = 3\,r_\mathrm{g}$. We fix
    the inclination at $i = 30^\circ$ and the outer edge of the disc
    at $r_\mathrm{out} = 400\,\rg$. Note that we always assume that
    the inner edge of the accretion disc coincides with the ISCO. All
    profiles are normalised to have equal area, i.e., the number of
    photons producing each reflection feature is equal.}
\end{figure*}
With the \textsc{relline\_lp} model it is possible to calculate the
predicted line shapes of broad emission lines for the different
parameters determining the setup in the lamp post geometry.
Figure~\ref{fig:lines_relline_lp} shows that the line shape is very
sensitive for certain parameter combinations and almost independent in
other cases.

The line shape is highly sensitive to a change in height of the
primary source (Fig.~\ref{fig:lines_relline_lp}a and~b). Especially
when assuming a rapidly rotating black hole, the line shape
dramatically changes from a really broad and redshifted line to a
narrow and double-peaked structure when increasing the height of the
source (Fig.~\ref{fig:lines_relline_lp}a). The same behaviour can be
observed for a negatively rotating black hole
(Fig.~\ref{fig:lines_relline_lp}b), but here differences are not as
large as in the previous case. Broad lines seen from a configuration
of a low primary source and a highly rotating black hole are also
sensitive to the photon index $\Gamma$ of the incident spectrum
(Fig.~\ref{fig:lines_relline_lp}c). In this case the line shape gets
broader for a softer incident spectrum. When fixing the height of the
irradiating source (Fig.~\ref{fig:lines_relline_lp}d--f), for a low
source height the shape is still sensitive to the spin
(Fig.~\ref{fig:lines_relline_lp}d). But already for a medium height of
25\,\rg the line shapes virtually coincide for all possible values of
black hole spin. In this case, even for a rapidly rotating black hole,
the line shape does not depend on the steepness of incident spectrum
(Fig.~\ref{fig:lines_relline_lp}f).

Using the \textsc{relline\_lp}-code, we are also able to compare line
shapes for moving (Sect.~\ref{sec:moving-jet-base}), elongated
(Sect.~\ref{sec:irrad-an-extend}), and accelerating
(Sect.~\ref{sec:jet-with-constant}) primary sources.  A change in
velocity only alters the line shape if the source is at low height and
the black hole is rapidly rotating
(Fig.~\ref{fig:lines_relline_lp}g). Changing either of these to
negatively rotating (Fig.~\ref{fig:lines_relline_lp}h) or a larger
source height (Fig.~\ref{fig:lines_relline_lp}i), different velocities
of the irradiating source only have a marginal effect on the line
shape. Similarly, measuring the radial extent of the primary source is
also not possible for all parameter combinations. Firstly, the
restrictions of measuring $a$ and $\Gamma$ for larger heights, as seen
in Fig.~\ref{fig:lines_relline_lp}e and~f, also apply here. Therefore
we fix these two values at $a=0.99$ and $\Gamma=2.0$. Setting the base
of the source at 3\,\rg\ and altering its height
(Fig.~\ref{fig:lines_relline_lp}j) does indeed result in great changes
in the line shape. These changes are very similar to a change in
height of a point-like primary source (see
Fig.~\ref{fig:lines_relline_lp}a). Hence, an elongated jet produces a
reflection feature similar to a point-like source with an effective
height. A similar behaviour can be observed when changing the base of
the primary source and leave the upper boundary constant at
100\,\rg. However, as the large upper part now dominates the
irradiation of the accretion disc, the profile is not very sensitive
to the location of the base of the source.  Finally, an accelerating
jet (Fig.~\ref{fig:lines_relline_lp}l) only influences the line shape
if already relatively high velocities are obtained at a height of
100\,\rg\ ($v_{100} > 0.99c$), as in the upper part of the jet more
and more photons are beamed away from the accretion disc due to the
highly relativistic movement of the emitting medium. The line profiles
for a strong acceleration thus resemble very closely the ones for a
lower top of the emitting source (Fig.~\ref{fig:lines_relline_lp}j),
which means that in reality we will not be able to measure if the
emitting source is accelerating without knowing the full geometry of
such a primary source.

In some sources with a low mass accretion rate, the disc might be
truncated further away from the black hole than the ISCO (see
\citealt{Esin1997a} and, e.g., the observations by
\citealt{Markowitz2009a,Svoboda2010a}). Generally, relativistic
emission lines from such truncated discs are even narrower than for
non or negatively spinning black holes. However, such lines can also
be explained by the irradiation from an largely elevated
\citep{Chiang2012a} or elongated source
(Fig.~\ref{fig:lines_relline_lp}k).

Besides the spin, the inclination of the system has also a strong
effect on the line shape \citep[see, e.g., ][]{Dauser2010a}. As the
inclination is mainly determined by defining the maximal extent of the
line at the blue side (Fig.~\ref{fig:lines_relline_lp}), the steep
drop is always at the same location for a fixed inclination and can
therefore be determined almost independently of the geometry.

In summary, Fig.~\ref{fig:lines_relline_lp} confirms that the
relativistic reflection feature is sensitive to very different
parameters in the lamp post geometry. However, as soon as the primary
source is not very close to the black hole or is elongated in the
radial direction, the dependency on parameters such as the spin of the
black hole or the incident spectrum is not large.

\subsection{Implications for Measuring the Spin of a Black Hole}
\label{sec:impic-meas-spin}

If we are interested in measuring the spin of the black hole by
analysing the relativistic reflection, Fig.~\ref{fig:lines_relline_lp}
and the discussion in the previous section help us to decide under
which conditions we are capable in doing so. Generally, these
conditions can be separated into two classes: Either the primary
source is compact and very close to the black hole, or the irradiating
emission comes from larger heights or an elongated structure. As has
been shown in the previous section, the latter case, a large height of
the primary source and an elongated structure, produces a very similar
irradiation of the accretion disc and therefore we use following
picture for this discussion: The base of the primary source is always
located at 3\,\rg\ and simply the top height of the elongated
structure is allowed to change. For simplicity we call it a ``compact
jet'' if the top height of the source is close to 3\,\rg\ and
``extended jet'' for larger values of the top height.

\begin{figure}
  \centering
  \includegraphics[width=\columnwidth]{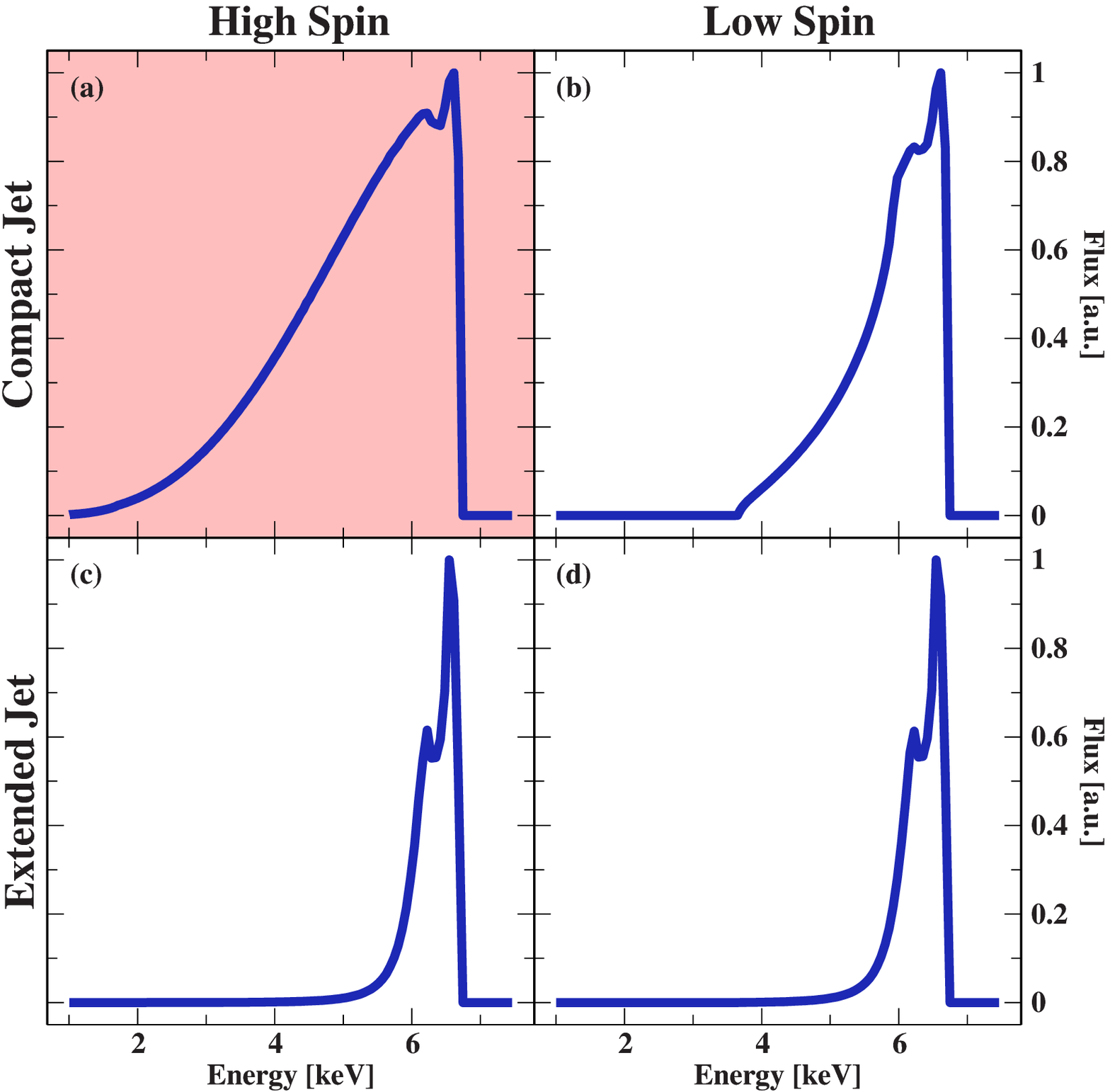}
  \caption{The line shape for combinations of high and low spin
    ($a=0.99$ and $a=0$) of the black hole and for irradiation by a
    compact jet (point source at $3\,\rg$) and an extended jet (from
    $3\,\rg$ to $100\,\rg$). Only for high spin and a compact, low jet
    the line will be observed as being broad (a, red field).}
  \label{fig:diagram}
\end{figure}
As an example, Fig.~\ref{fig:diagram} shows how a line profile for
different spin would look like for a compact and extended jet. In the
case of a compact jet the sensitivity on the spin is high (see
Fig.~\ref{fig:lines_relline_lp}d) and therefore we can clearly see the
difference of a really broad line for high spin
(Fig.~\ref{fig:diagram}a) compared to a much narrower line for low
spin (Fig.~\ref{fig:diagram}b). If the irradiating source is
elongated, the line profiles for high and low spin are very similar
and especially really narrow (Fig.~\ref{fig:diagram}c,d). At first
this underlines what was already discussed above in the context of
Fig.~\ref{fig:lines_relline_lp}: If we have an elongated source we we
will not be able to constrain the spin at all.

But the diagram in Fig.~\ref{fig:diagram} reveals a much larger
problem we have to deal with. Namely three of the four lines in the
Figure will be detected as narrow lines and when fitted with any model
simulating broad lines, will result in very low spin values. But this
would be wrong in the case of Fig.~\ref{fig:diagram}c, where the black
hole is rapidly rotating, but only the elongated structure renders the
line profile narrow. Reversing the arguments, this implies that
observations of a narrow line\footnote{narrow, but still broad in CCD
  resolution} do not allow to draw any conclusions about the spin of
the black hole without knowing the exact geometry of the primary
source. Hence, any black hole which was inferred to be slowly rotating
due to its narrow reflection feature might very well be rapidly
spinning if the accretion disc is irradiated by a source which is not
compact or close to the black hole. This issue makes it even harder to
produce any statistics on the spin of a sample of AGN by reflection
measurements, as only from a sub-sample of the sources, the ones with
relatively broad reflection features, the spin can be determined
reliably.

\subsection{Simulated Observations with current Instruments}
\label{sec:simul-observ-with}

We performed several simulations in order to quantify the
argumentation of the previous section, on how sensitive spin
measurements can be depending on the geometry of the irradiating
source. For this purpose we use the EPIC-pn camera
\citep{Strueder2001a} of the \textsl{XMM-Newton} satellite
\citep{Jansen2001a}, a widely used instrument for measuring reflection
features and determining the spin of a black hole \citep[see,
e.g.,][]{wilms2001a,Fabian2009a,Duro2011a}.

\begin{figure*}
  \centering
  \includegraphics[width=\textwidth]{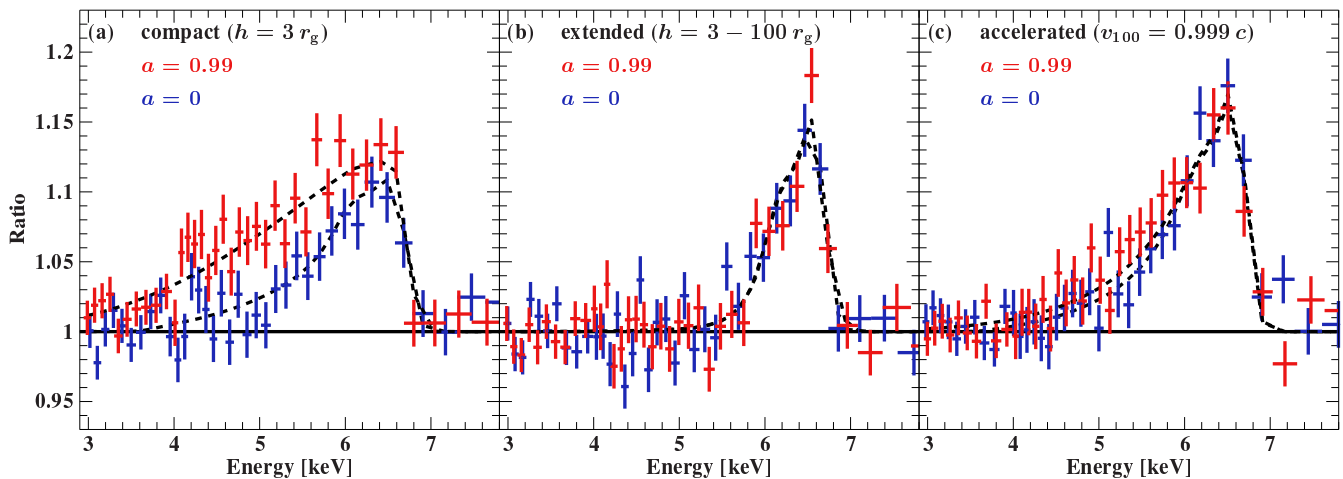}
  \caption{A simulated \textsl{XMM-Newton} observation of $100\,$ksec
    exposure of the Fe K$\alpha$ line for (a) a compact ($h =
    3\,\rg$), (b) an extended jet ($h = 3$--$100\,\rg$), and (c) again
    an extended jet ($h = 3$--$100\,\rg$), but now extremely
    accelerated such that $v_{100} = 0.999\,c$. Parameters were taken
    to be similar to MCG$-$6-30-15 in the \textsl{XMM-Newton}
    observation with Obs.~ID 0111570101 \citep{wilms2001a}. In order
    to obtain consistent line strengths we fixed the desired geometry
    and spin value and fitted the normalisation of the line to the
    data.  The equivalent widths for high spin compared to
    non-spinning were determined to 432\,eV vs.\ 175\,eV (a), 110\,eV
    vs.\ 109\,eV (b), and 255\,eV vs.\ 208\,eV (c). }
  \label{fig:sim_diagram}
\end{figure*}
Figure~\ref{fig:sim_diagram} shows 100\,ksec observations of an iron
K$\alpha$ feature for different cases, namely of a compact primary
source (Fig.~\ref{fig:sim_diagram}a) and an elongated source
(Fig.~\ref{fig:sim_diagram}b), which might also be accelerating
(Fig.~\ref{fig:sim_diagram}c). Looking at the equivalent widths in the
figure clearly reveals by what amount the line changes from a highly
to a non rotating black hole. For the canonical case
(Fig.~\ref{fig:sim_diagram}a), a compact primary source, there are
large differences visible, already for a very common exposure time of
100\,ksec. On the other hand, in the case of an elongated source
(Fig.~\ref{fig:sim_diagram}b), there is virtually no difference at all
and even future instruments will not be able to deduce the spin in
this case. If the elongated source is accelerating strongly
(Fig.~\ref{fig:sim_diagram}c), slight differences arise (see dashed
lines) but still stay deeply hidden within the uncertainties of the
measurement.

As illustrated in Sect.~\ref{sec:impic-meas-spin}, the inability of
measuring the spin for certain geometries is not the major problem.
This approach assumes that we know the geometry of the primary source,
which is generally not the case. Hence, the interesting question is,
what spin value would we measure when we assume the ``standard''
geometry ($I \propto r^{-3}$) currently used?
\begin{table}
\centering 
\caption{This table summarises the result of a simulation which tries
  to illustrate what spin value would be obtained if we use the
  ``standard'' geometry (i.e., where the emissivity is proportional to
  $r^{-3}$) for describing data, which was originally produced by a more
  complex geometry of the primary source (compact, extended, or
  accelerated). This simulation was performed for two different values
  of spin ($a=0.99$ or $a=0.0$). All details are identical to the
  simulation described in Fig.~\ref{fig:sim_diagram}.  }
\include{sim_conf_table}
\label{tab:sim_conf_table}
\end{table}
Table~\ref{tab:sim_conf_table} summarises the results of such an
investigation. As expected, simulating a compact source yields nicely
consistent results of spin value when fitted by the standard geometry.
Contrary to that the elongated source, which produces always a very
narrow line (see Fig.~\ref{fig:sim_diagram}b), is always fitted by a
very low spin, independent of the actual spin of the black hole.
Especially for the highly rotating black hole this solution is in
great contrast to the original value and even for a 100\,ksec
observation with current instruments far outside the confidence
limits. If the elongated source now strongly accelerates while
emitting photons, the obtained spin values resemble more the true ones
($a^*=0.5^{+0.2}_{-0.3}$ for $a_\mathrm{bh}=0.99$ and
$a^*=-0.3^{+0.4}_{-0.7}$ for $a_\mathrm{bh}=0.0$). However, these
values still do not agree with the actual spin of the black hole. As
this misleading determination of the spin does not fail due to low
sensitivity but a lack of information of the primary source, future
satellite missions or longer exposures will not be able to solve this
issue.

\section{Summary and Conclusions}
\label{sec:summary}

In this paper we introduced a consistent approach in general
relativity for modelling the irradiation of an accretion disc in the
lamp post geometry. Besides a stationary point source, we also
showed how a moving, a radially elongated, or an accelerating
primary source can be easily modelled by slightly extending the
formalism. We presented an extension to the \textsc{relline} code
\citep{Dauser2010a}, which can be used for model fitting in common
data analysis programs and is capable of describing all different
kinds of primary sources mentioned above. We could show that only for
a compact primary source the spin of the black can be measured
properly. Elongated sources will always produce a narrow relativistic
reflection feature. Even if we introduce a strong acceleration, the
simulation showed that the spin cannot be determined for such a setup
and might be confused with a compact irradiating source around a non-
or negatively rotating black hole.

For the different geometries of the primary source we analysed the
difference in emissivity profiles. Most importantly we could see that
the innermost part (i.e., $r<2\,\rg$), which is also the steepest part
of the profile, is completely independent of the geometry and movement
of the source \citep[see also ][]{Fukumura2007a}. Hence, by changing
the geometry in our lamp post setup, we will not be able to create
steeper profiles than the standard one for a point source in order to
explain the really steep emissivity indices measured in many
observations \citep[see, e.g.,
][]{wilms2001a,fabian2002a,Fabian2004a,Brenneman2006a,Dauser2012a}.
Even sources extended perpendicular to the rotational axis seem not to
be able to alter the steepness of the emissivity \citep{Wilkins2012a}.
However, these steep emissivities might be artificially created due to
an over-simplified modelling. Recent analyses show that describing
the reflection at the accretion disc with a constant ionisation,
instead of an ionisation gradient that would be expected from basic
arguments \citep[using, e.g.,][]{Shakura1973a}, would result in
over-estimating the emissivity index \citep{Svoboda2012a}. Moreover we
showed that this independence of the inner part of the emissivity
profile from the geometry comes from the fact that this steepening is
mainly due to the strong blue-shift the photons exhibit during their
flight towards the accretion disc. Although the strong focusing of the
photons close to the black hole is commonly used to explain this
steepening, Fig.~\ref{fig:simple_gamma_eshift} evidently showed that
this effect actually only contributes a minor part to the steepness
compared to the energy shift.

Using the new \textsc{relline\_lp} model, a variety of broad emission
lines for very different parameter combinations was produced. This
showed that only compact sources close to the black hole produce the
really broad lines and are sensitive to parameters like the black hole
spin or the steepness of the incident spectrum. Hence, observing a
broad line already strongly constrains the primary emission region to
be very close to the black hole, which also implies that most of the
reflected radiation originates from the innermost parts of the
accretion disc. Moreover the narrow line profile created by an
elongated primary source is very similar and therefore hardly
distinguishable from a point source at an effective, intermediate
height. Introducing an acceleration of the elongated source
effectively reduces the flux from the upper part of the source. This
acceleration leads to a broadening of the line shape, which means that
it becomes again more sensitive to the black hole parameters and more
similar to the line profile created by a compact source at low height.

Simulating observations with current instruments, the line profile of
elongated sources will always be rather narrow. Applying standard line
models to X-ray spectra of such systems will always result in a low
spin of the black hole, although this might not be the actual case. We
therefore conclude that only a really broad emission line will give us
the possibility to determine the spin of the black hole unambiguously.
Such systems must have a compact emitting source and a rapidly
rotating black hole. In all other cases with narrower lines, the
emission line can either be described by a slower rotating black hole
or an elongated source, which could have any arbitrary spin. The
thought that a narrow, relativistic reflection feature might be
produced by a primary source at a larger distance to the black hole
not new. \citet{Vaughan2004b} suggested such a geometry to explain the
reflection feature in \mbox{Akn~120}, as it required a flat emissivity
at the inner edge of the accretion disc. In agreement with our results
presented above, \citet{Chiang2012a} conclude for
\mbox{XTE~J1652$-$453} that the spin cannot be constrained if the
primary source is assumed to be located at a large height above the
black hole.

We note that in some sources broad lines have been seen to be
changing. Galactic black holes are seen in two states, a hard state
where the X-ray continuum is a power-law with an exponential cutoff
that can be well described by a Comptonization continuum with
(relativistically smeared) reflection, and a soft state in which
thermal emission from an accretion disc dominates the emission
\citep[see, e.g., ][for a review]{McClintock2006b}. During the hard
state a relativistic outflow is present in these systems, which is
quenched during the soft state \citep{Fender2004a}. This jet can carry
a significant percentage of the total accreted power \citep[][and
references therein]{Russell2007a}. In many models for the black hole
states it is assumed that the inner disc radius is receded outwards
during the hard state and that it only extends to the ISCO during the
soft state \citep[e.g.,][]{Esin1997a,Dove1997a,Fender2004a}. If the
Comptonizing plasma is located close to the accretion disc, an inward
moving accretion disc would increase Compton cooling of the
Comptonizing plasma, which then collapses, resulting in the soft
state. One of the major observational arguments for this picture are
changes of the relativistically broadened line in these systems. For
example, the Galactic black hole GX~339$-$4 is seen to have a broad
line during the more luminous phases of the hard state, i.e., closer
to the soft state \citep[][but note
\citealt{Kolehmainen2011a}]{Miller2008a} and a narrow line during the
fainter phases of the hard state \citep[][see also
\citealt{Nowak2002a}]{Tomsick2008a}. While this behaviour is consistent
with a receding disc, given the discussion above this picture is not
as clear cut anymore. The broad band radio to X-ray spectra of
galactic black holes are also consistent with emission from the base
of a jet, with a statistical quality comparable to that to the
disc-corona models \citep{Markoff2005a,Nowak2011a}. Furthermore, the
jet properties in these sources are observed to change as the state
approaches the intermediate and soft states, including the observation
of single radio blob ejections close to transitional states
\citep{Fender2006a,Wilms2007a,Zdziarski2011a}. If this is the case,
then according to the discussion above the change in line width from
narrow to broad seen in GX~339$-$4 could also be due to a change in
the jet properties from an extended jet far away from the soft state
towards a compact jet, rather than a change in the inner disc radius
\citep[see also][]{Miller2012a}. 

For observations of sources where the line shape changes only on very
long timescales, such as AGN, our results imply that solely relying on
spectral fits we are only able to measure the spin of rapidly rotating
black holes. All other spin values obtained from fitting reflection
spectra are not trustworthy, as they can be easily explained by
changing the geometry of the irradiating source. This result is of
significant importance, e.g., when trying to use spectral black hole
spin measurements to infer the spin distribution of the local black
hole population, as it introduces a significant bias towards sources
with small jet sizes.

Fortunately, however, reflection measurements are only one of several
independent ways to determine black hole spins. For Galactic sources,
where the accretion disc temperature is large enough that it can be
observed in the soft X-rays, it is possible to use the shape of the
accretion disc spectrum to infer the black hole spin \citep[][but see
\citealt{Dexter2012a}]{McCLintock2010a}. Where a broad line is
observed, the results of continuum spin measurements and line
measurements agree (e.g., \citealt{Gou2011a} and \citealt{Duro2011a}
for the case of Cyg~X-1). Furthermore, X-ray reverberation techniques
allow constraining the location of the primary source through
measuring X-ray time lags between the continuum and the reflection
spectrum \citep[][and references
therein]{Stella1990a,Matt1992b,Reynolds1999a,poutanen2002a,Uttley2011a}.
Despite the comparatively small effective area of current X-ray
instruments, such measurements have already been successfully applied
to a small subset of extragalactic black holes, e.g., 1H0707$-$495
\citep{Fabian2009a,Zoghbi2010a,Zoghbi2011a,Kara2012a}, NGC~4151
\citep{Zoghbi2012a}, MCG$-$6-30-15 and Mrk~766
\citep{Emmanoulopoulos2011a}, and PG~1211$+$143
\citep{Marco2011a}). In all these measurements the primary source was
constrained to be compact, extending only in the order of a
few-to-tens $\rg$ away from the black hole. Current concepts for the
next generation of X-ray satellites such as the Large area Observatory
for X-ray Timing \citep[\textsl{LOFT};][]{Feroci2012a} or the Advanced
Telescope for High Energy Astrophysics
\citep[\textsl{ATHENA};][]{Barcons2012a} all have effective areas of
several square meters. With such large collecting areas it will be
possible to measure high-quality X-ray spectra for a larger sample of
AGN within the typical light travel timescales for the lamp post
model. These facilities will allow X-ray reverberation measurements to
constrain the emission geometry of the observed sources such that the
degeneracies pointed out above can be broken for a larger sample of
sources.

\emph{Acknowledgements.} We acknowledge support from the European
Commission under contract ITN 215212 ``Black Hole Universe'', by a
fellowship from the Elitenetzwerk Bayern, and by the Deutsches Zentrum
f\"ur Luft- und Raumfahrt under contract number 50 OR 1113. We thank
John E.~Davis for the development of the \textsc{SLxfig} module used
to prepare the figures in this paper, Katja Pottschmidt and John
Tomsick for useful discussions, and the referee, Andy Fabian, for his
constructive comments which helped improving this paper.

\bibliographystyle{mn2e_williams} \bibliography{mnemonic,mn_abbrv,bib}

\appendix
\label{lastpage}

\end{document}

%% file: sim_conf_table.tex
 \begin{tabular}{cccc} 
\hline\hline  & compact & extended & accelerated \\ \hline
 $a_\mathrm{bh}=0.99$ & $1.00^{+0.00}_{-0.06}$ & $-1.00^{+0.27}_{-0.00}$ & $0.45^{+0.24}_{-0.34}$ \\
 $a_\mathrm{bh}=0.0$ & $0.2^{+0.5}_{-0.6}$  & $-1.0^{+0.4}_{-0.0}$  & $-0.3^{+0.4}_{-0.7}$  \\
\hline\hline\end{tabular}